\documentstyle[12pt]{article}

\def\al{\alpha}
\def\be{\beta}
\def\ga{\gamma}

\def\et{\eta}

\def\ka{\kappa}
\def\la{\lambda}

\def\rh{\rho}

\def\si{\sigma}

\def\ta{\tau}

\def\ch{\chi}

\def\om{\omega}
\def\Ga{\Gamma}

\def\Ph{\Phi}

\def\cl{{\cal L}}

\def\cE{{\cal E}}
\def\cL{{\cal L}}

\def\fr#1#2{{{#1} \over {#2}}}

\def\pt#1{\phantom{#1}}
\def\vev#1{\langle {#1}\rangle}

\def\half{{\textstyle{1\over 2}}}
\def\quar{{\textstyle{1\over 4}}}
\def\frac#1#2{{\textstyle{{#1}\over {#2}}}}

\def\lsim{\mathrel{\rlap{\lower4pt\hbox{\hskip1pt$\sim$}}
    \raise1pt\hbox{$<$}}}
\def\gsim{\mathrel{\rlap{\lower4pt\hbox{\hskip1pt$\sim$}}
    \raise1pt\hbox{$>$}}}
\def\sqr#1#2{{\vcenter{\vbox{\hrule height.#2pt
         \hbox{\vrule width.#2pt height#1pt \kern#1pt
         \vrule width.#2pt}
         \hrule height.#2pt}}}}

\def\lsc#1#2#3{\om_{#1#2#3}}

\def\lulsc#1#2#3{\om_{#1\pt{#2}#3}^{{\pt{#1}}#2}}

\def\vb#1#2{e_{#1}^{{\pt{#1}}#2}}

\def\lvb#1#2{e_{#1#2}}

\def\lsc#1#2#3{\om_{#1#2#3}}

\def\lulsc#1#2#3{\om_{#1\pt{#2}#3}^{{\pt{#1}}#2}}

\newcommand{\beq}{\begin{equation}}
\newcommand{\eeq}{\end{equation}}
\newcommand{\bea}{\begin{eqnarray}}
\newcommand{\eea}{\end{eqnarray}}
\newcommand{\rf}[1]{(\ref{#1})}

\begin{document}

\begin{center}
{\bf Effects of Spontaneous Lorentz Violation in Gravity
\footnote{\tenrm
Talk presented at the meeting, 
From Quantum to Emergent Gravity: Theory and Phenomenology,
June 2007, Trieste, Italy.}
\\}
\end{center}

\begin{center}
{Robert Bluhm \\ }
{Physics Department \\ }
{Colby College \\ }
{Waterville, ME 04901 USA\\}
\end{center}

\begin{abstract}
\noindent
Spontaneous breaking of local Lorentz symmetry 
occurs when a local vector or tensor field acquires a 
nonzero vacuum expectation value.  
The effects of such breaking are examined in the context of gravity theory.  
These include an associated spontaneous breaking of 
diffeomorphism symmetry and generation of massless 
Nambu-Goldstone modes.  
The possibility of a Higgs mechanism is examined as well, 
and it is found that the conventional Higgs mechanism 
(giving rise to massive gauge fields) does not occur in a Riemann spacetime.  
However, in a Riemann-Cartan spacetime a Higgs mechanism 
involving the spin connection is possible.  
Despite the lack of a conventional Higgs mechanism in Riemann spacetime, 
additional massive modes  involving the metric can appear 
through unconventional processes that have no analogue in nonabelian gauge theory.  
The effects of these types of processes are illustrated using a specific model, 
known as a bumblebee model, in which a vector field acquires a vacuum value.
\end{abstract}

\section{Introduction}

Interest in the possibility of Lorentz violation has increased
significantly over the years, 
as it has been realized that effects arising
at the Planck scale might give rise to small breakings of Lorentz 
symmetry at low energy.
These include mechanisms in string theory, 
field theories in noncommutative geometry, 
quantum gravity, effects due to modified dispersion relations, etc.
(For reviews, see \cite{cptmeetings,Matreview,SRreview}).
At the same time,
a comprehensive phenomenological investigation of
Lorentz violation has been initiated
that has led to a number of new high-precision tests of Lorentz invariance.
This was facilitated by development of the Standard-Model Extension (SME)
\cite{sme,RBsme},
which provides the most general effective field-theoretical framework 
incorporating Lorentz and CPT violation.
Using the SME, 
detailed investigations of Lorentz breaking can
be conducted in the context of high-energy particle physics,
gravitational physics, nuclear and atomic physics,
and astrophysics.
This on-going effort has pushed experimental bounds on some
forms of Lorentz breaking
well beyond levels associated with suppression by the Planck mass.
Nonetheless,
many signals remain untested.
In addition to these phenomenological investigations,
the idea of Lorentz breaking continues to undergo theoretical scrutiny.
In particular,
possible effects of Lorentz breaking in the context of gravity have begun to be explored,
and in certain cases these ideas can lead to interesting prospects for alternative explanations of
such things as dark matter and dark energy.

One of the more elegant ideas for Lorentz violation 
is that this symmetry might be spontaneously broken
\cite{ks}.
Indeed, it was the idea that mechanisms in string field theory
might lead to spontaneous Lorentz violation 
that helped stimulate
much of the current interest in the topic of Lorentz breaking.
Moreover,
one of the primary interpretations of the coefficients in the SME is that they
are vacuum expectation values (combined with Yukawa couplings)
of tensor fields that couple to conventional matter at low energy.
Thus, one of the products of a comprehensive investigation of
Lorentz violation using the SME is a survey for possible signals of 
spontaneous Lorentz breaking originating from the Planck scale.

However,
as soon as one begins to discuss the idea of spontaneous
symmetry breaking,
well known results from particle physics immediately come into play.
These include the possible appearance of massless 
Nambu-Goldstone (NG) modes,
the possibility of a Higgs mechanism,
and the question of whether additional massive modes 
(analogous to the Higgs boson) can arise.

It is these types of effects that are examined here for the case where
it is local Lorentz symmetry that is spontaneously broken.
Clearly, any processes generating massless or massive modes
can have important implications for phenomenology.
This is particularly the case for spontaneous breaking of
Lorentz symmetry in the context of gravity,
where the effects of the NG and massive modes might influence
gravitational propagation or alter the form of the static Newtonian potential.
Thus, in addition to looking at the fate of the NG modes in general and the
question of whether a Higgs mechanism can occur,
it is important as well to look at the role of the NG and massive modes in
the context of specific models that permit an examination of their
effects on gravity.
The simplest example is for the case of a vector field,
where such models are known as bumblebee models
\cite{ks,akgrav}.
It is this type of model that is used here to illustrate 
the effects of the NG and massive modes.
This is then followed by a more general discussion of
phenomenology.

Many of the main results presented here,
including background on bumblebee models,  
are described in greater detail in 
\cite{ks,akgrav,rbak,rbffak},
as well as in the references cited within these works.

\section{Spontaneous Lorentz Breaking}

In special relativity, Lorentz symmetry is a global symmetry
consisting of rotations and boosts
However, in curved spacetime, in a gravitational theory, 
Lorentz symmetry is a local symmetry. 
It transforms local vectors and tensors in the tangent
plane at each spacetime point.
In addition to being locally Lorentz invariant, a gravitational theory is also
invariant under diffeomorphisms.  
These transformations act on tensor and vector fields defined on
the spacetime manifold.
Typically, it is the diffeomorphism symmetry that is more readily
apparent in a gravitational theory,
since local Lorentz symmetry acts only in local frames.
However, ultimately both types of transformations are important,
and a complete discussion of Lorentz breaking in gravity theory
should include an examination of its effects on diffeomorphisms as well.

One way to reveal the transformation properties of vectors and
tensors under both local Lorentz transformations and
diffeomorphisms is by using a vierbein formalism.
The vierbein $e_\mu^{\,\,\, a}$
relates tensor components defined with respect to a local basis, 
e.g., $T_{abc}$
(where Latin indices denote components with respect to a local frame),
to those defined with respect to the spacetime coordinate system,
e.g., $T_{\la\mu\nu}$
(where Greek indices label the spacetime frame) .
For example,
the spacetime metric and local Minkowski metric are related by
\begin{equation}
g_{\mu\nu} = e_\mu^{\,\,\, a} e_\nu^{\,\,\, b} \eta_{ab}  .
\label{vier}
\end{equation}       
Similarly,
for an arbitrary tensor,
\begin{equation}
T_{\lambda\mu\nu} \, = e_\lambda^{\,\,\, a} e_\mu^{\,\,\, b} e_\nu^{\,\,\, c} \, T_{abc} .
\label{T}
\end{equation}       
A vierbein formalism also allows spinors to be incorporated into the theory,
and it naturally parallels gauge theory,
with Lorentz symmetry and diffeomorphisms both
acting as local symmetry groups.
In a vierbein formalism,
the spin connection $\omega_\mu^{\,\,\, ab}$ enters in covariant
derivatives that act on local tensor components and plays the role of
the gauge field for the Lorentz symmetry.
In contrast,
the metric excitations act as the gauge
fields for the diffeomorphism symmetry.
When working with a vierbein formalism,
there are primarily two geometries that can be distinguished.
In Riemannian geometry (with no torsion), 
the spin connection is nondynamical.  
It is purely an auxiliary field that does not propagate.  
However, in Riemann-Cartan geometry (with nonzero torsion), 
the spin connection must be treated as independent degrees of freedom
that in principle can propagate.

In considering theories with violation of Lorentz and
diffeomorphism symmetry it is important to distinguish
between {\it observer} and {\it particle} transformations
\cite{sme}.
Under an observer general coordinate or local
Lorentz transformation,
vectors and tensors remain unchanged,
while the coordinate bases used to define their components transform.
In contrast,
particle diffeomorphisms and Lorentz transformations 
change vectors and tensors,
while leaving unchanged the coordinate systems and basis vectors.
In theories with no symmetry breaking,
the transformation laws for observer and particle transformations
are inversely related but otherwise are similar in form.
However,
if the symmetries are spontaneously broken
and fields are divided into vacuum values and excitations,
the transformation laws for these will differ for
the observer and particle transformations.

A fundamental premise is that a physical theory should always be observer independent.  
This includes even when Lorentz symmetry and diffeomorphisms
are either explicitly or spontaneously broken.
In fact,
the SME is based on this.
It is formulated as a lagrangian-based field theory that
is fully invariant under observer
general coordinate transformations
and local Lorentz transformations.

In a theory with spontaneous breaking of a particle spacetime symmetry,
the lagrangian still remains invariant under the broken symmetry,
and the full equations of motion remain covariant.
However,
fixed vacuum-valued fields appear that cannot be transformed
under the particle transformations.
Interaction terms involving these fixed background vacuum 
fields also appear in the equations of motion,
which by themselves break the particle symmetry.
It is the interaction with these vacuum fields that 
can lead to physical effects of the broken particle symmetry
that can be tested in experiments.

In a gravitational theory,
local Lorentz symmetry is spontaneously broken when a local tensor 
field acquires a vacuum expectation value (vev).
For example,
for the case of a three-component tensor,
\begin{equation}
<T_{abc}> \, = t_{abc} .
\label{Tvev}
\end{equation}       
The vacuum of the theory then has preferred spacetime directions
in the local frames, 
which spontaneously breaks the particle Lorentz symmetry.

Spontaneous Lorentz breaking can be introduced into a theory
dynamically by adding a potential term $V$ to the Lagrangian.
For example, a potential of the form
\beq
V \sim(T_{\la\mu\nu} \,
g^{\la\al} g^{\mu\be} g^{\nu\ga}  \,
T_{\al\be\ga}
\pm \, t^2)^2 ,
\label{VT2}
\eeq
consisting of a quadratic function of products of the tensor
components $T_{\la\mu\nu}$,
has a minimum when \beq
T_{\la\mu\nu} \, g^{\la\al} g^{\mu\be} g^{\nu\ga} \,
T_{\al\be\ga} = \mp \, t^2 .
\label{condT}
\eeq
Note that the sign on the right-hand side depends on the timelike 
or spacelike nature of the tensor components.
Solutions of Eq.\ \rf{condT} span a degenerate space of
possible vacuum solutions.
Spontaneous Lorentz breaking occurs when a
particular vacuum value $t_{abc}$ in the local frame is chosen,
satisfying the condition
\beq
\mp t^2 =
t_{abc} \,
\et^{pa} \et^{qb} \et^{rc} \,
t_{pqr} .
\eeq
Alternatively,
a potential with a Lagrange multipler field $\la$
can impose \rf{condT} directly as a constraint,
which also leads to spontaneous selection of a vacuum value $t_{abc}$.

\section{Nambu-Goldstone Modes}

To examine the fate of the NG modes in a theory with spontaneous
Lorentz violation,
a general approach can first be followed.
Consider a theory with a tensor that has a nonzero
vev in a local Lorentz frame,
for example,
$<T_{abc}> \, = t_{abc}$.    
Such a vev spontaneously breaks particle local Lorentz symmetry.  
In addition,
the vierbein also has a constant or fixed background value.
For example, 
in a background Minkowski spacetime,
\begin{equation}
<e_\mu^{\,\,\,\, a}> \, = \delta_\mu^{\,\,\, a} .
\label{evev}
\end{equation}       
The spacetime tensor therefore has a vev as well,
\begin{equation}
<T_{\lambda\mu\nu}> \, = t_{\lambda\mu\nu} ,
\label{Tmunuvev}
\end{equation}     
which is obtained when $<e_\mu^{\,\,\,\, a}>$ acts on $t_{abc}$.
This fixed vacuum value for $T_{\lambda\mu\nu}$means 
that particle diffeomorphisms are spontaneously broken.
Thus, a first general result is that
spontaneous breaking of local Lorentz symmetry
implies spontaneous breaking of diffeomorphisms.

Spontaneous breaking of these symmetries implies that NG
modes should appear (in the absence of a Higgs mechanism).
This raises the question of how many NG modes can appear. 
The usual rule is that there can be up to as many 
NG modes as there are broken symmetries. 
In this case,
maximal symmetry-breaking would yield six broken Lorentz generators 
and four broken diffeomorphisms.
Therefore, there can be up to ten NG modes in general.

A related question asks where the ten NG modes reside.
In general, the answer depends on the choices of gauge.  
However, one natural choice is to put all ten NG modes into the vierbein,
as a simple counting argument shows is possible.
The vierbein $e_\mu^{\,\,\, a}$ has 16 components.  
With no spontaneous Lorentz breaking, 
the six Lorentz and four diffeomorphism degrees
of freedom can be used to reduce the vierbein down to six independent degrees
of freedom.  
(Note that a general gravitational theory can have six propagating metric modes; 
however, general relativity is special in that there are only two). 
In contrast, in a theory with spontaneous
Lorentz breaking, 
where all ten spacetime symmetries are broken,
the vierbein can have 16 propagating degrees of freedom.
Therefore, a second result is that
in a theory with spontaneous Lorentz breaking,
up to ten NG modes can appear and all of them can naturally
be incorporated as degrees of freedom in the vierbein.

These results can be obtained as well using an expansion of  
the vierbein in terms of infinitesimal excitations about the vacuum.
For such small excitations,
the distinction between local and spacetime components can
be dropped (with Greek letters being used for both from here on).
The vierbein (with lowered indices) is then written as
\beq
\lvb \mu \nu = \et_{\mu\nu} + (\half h_{\mu\nu} + \ch_{\mu\nu}) ,
\eeq
in terms of symmetric components,
$h_{\mu\nu} = h_{\nu\mu}$,
and antisymmetric components,
$\ch_{\mu\nu} = -\ch_{\nu\mu}$.
Next, consider small excitations of the tensor field about its vacuum value.
For the case of a three-component tensor,
these have the form
\beq
\ta_{\la\mu\nu} = (T_{\la\mu\nu} - t_{\la\mu\nu}) .
\eeq
The NG modes are the field excitations that stay within
the minimum of the potential $V$.
They therefore obey the condition \rf{condT}.
A solution of this condition is given by the vierbein
acting on the local vev, 
and is equal to
\beq
T_{\la\mu\nu} = \lvb \la \al \lvb \mu \be \lvb \nu \ga \, t^{\al\be\ga} .
\eeq
Inserting the expansion of the vierbein into this equation and solving
for the tensor-field excitations to lowest order gives
\beq
\ta_{\la\mu\nu} \approx
(\half h_{\la\al} + \ch_{\la\al}) 
t^\al_{\pt{\al}\mu\nu}
+ (\half h_{\mu\al} + \ch_{\mu\al}) 
t^{\pt{\la}\al}_{\la\pt{\al}\nu} 
+ (\half h_{\nu\al} + \ch_{\nu\al}) 
t^{\pt{\la\mu}\al}_{\la\mu} .
\eeq
Evidently,
it is the combination $(\half h_{\mu\nu} + \ch_{\mu\nu})$ that contains
the NG fields.
Indeed,
with the appropriate gauge choices,
an expansion of this combination as virtual local
Lorentz transformations and diffeomorphisms shows
explicitly that the NG modes reside in these components
\cite{rbak}.
In this form,
the Lorentz NG modes can be associated with the six antisymmetric
components $\ch_{\mu\nu}$ while the diffeomorphism NG modes
can be most closely associated with the four 
gauge degrees of freedom in $h_{\mu\nu} $.

\section{Gravitational Higgs Mechanism}

In the context of gravity,
since Lorentz symmetry is a local symmetry,
the possibility of a Higgs mechanism naturally arises.
However,
since there are two sets of broken symmetries (Lorentz and diffeomorphisms) 
there are potentially two associated  Higgs mechanisms. 
Furthermore,
there is also the possibility that additional massive modes can arise
as field excitations obeying $V^\prime \ne 0$ that do not stay in the
potential minimum.
This section discusses these possibilities.

First,
consider the case of diffeomorphisms.
For this symmetry, 
the corresponding gauge fields are the metric 
(or vierbein) excitations,
and therefore a conventional Higgs mechanism would presumably give
rise to mass terms for the metric.
However,
it was previously shown 
that the usual Higgs mechanism involving the metric does not occur
\cite{ks}.
This is because in the conventional Higgs mechanism
the quadratic terms (that give rise to mass terms)
come from the kinetic terms for the field acquiring a vev
(the tensor field in this case).
These kinetic terms consist of products of 
covariant derivatives acting on the tensor field.
However,
for diffeomorphism-covariant
(as opposed to gauge-covariant) derivatives
it is the connection that appears in quadratic form.
For example,
\beq
(D_\rh T_{\la\mu\nu})^2 \sim ( \Ga^\si_{\rh\la} t_{\si\mu\nu})^2 + \cdots .
\eeq
Since the connection consists of derivatives of the metric,
and not the metric itself,
there are no mass terms generated for the metric.
As a result,
there is no conventional Higgs mechanism for the metric.

However, it was also pointed out in Ref.\ \cite{ks}
that the form of the potential $V$, 
as for example in Eq.\ \rf{VT2}, 
does permit quadratic terms involving excitations of the metric to appear.
This results in an alternative form of the Higgs mechanism
that has no direct analogue in nonabelian gauge theory.
This is because in nonabelian gauge theory, 
the potential $V$ only involves the scalar Higgs fields
(and not the gauge fields).
However, in the case of spontaneous diffeomorphism breaking,
it is combinations of both the tensor field and the metric field excitations
that acquire quadratic mass terms.
Since the metric appears in these terms,
but not in the usual Fierz-Pauli form,
it becomes possible in principle
to generate mass terms that 
avoid the van Dam, Veltmann, and Zakharov discontinuity
\cite{vdvz}.
However,
the question of whether ghost modes are generated must
also be addressed.
This typically becomes a model-dependent issue,
since the form of the kinetic terms for the tensor fields 
can influence whether the massive-mode excitations propagate
and whether ghost modes appear.
In some models,
the massive modes do not propagate,
but instead remain auxiliary fields.
However, even in these cases,
the massive modes can have an influence on gravitational interactions,
including possible modifications of the Newtonian potential
\cite{rbffak}.

Summarizing for the case of diffeomorphisms,
the general results are that
there is no conventional Higgs mechanism for the graviton;  
however, mass terms involving the metric may 
arise due to the form of the potential $V$,
resulting in an alternative Higgs mechanism that has no
direct parallel in nonabelian gauge theory.

The next question is whether a Higgs mechanism can occur 
stemming from the broken local Lorentz symmetry.
For the Lorentz symmetry, 
it is found that a conventional Higgs mechanism can occur
\cite{rbak}.
The relevant gauge field for the Lorentz symmetry
is the spin connection.  
It appears directly in expressions for covariant derivatives 
acting on local tensor components.
For the local tensor that acquires a vev,
quadratic mass terms for the spin connection can be generated,
following the usual Higgs mechanism.
For example, kinetic terms of the form
\beq
(D_\rh T_{\la\mu\nu})^2 \sim ( \om_{\rh \pt{\al} \la}^{\pt{\la}\al} \,  t_{\al\mu\nu})^2 + \cdots
\eeq
can generate quadratic terms for the spin connection.
However, a viable Higgs mechanism of this form
involving the spin connection can 
only occur if the spin connection itself is a dynamical field.  
This requires that there is nonzero torsion and therefore
that the geometry is Riemann-Cartan.
Thus, a Higgs mechanism for the spin connection is possible, 
but only in a Riemann-Cartan geometry.
Constructing a ghost-free model with a propagating spin connection
is known to be a challenging problem
\cite{sc}.
Incorporating Lorentz violation may lead to the appearance of 
additional mass terms,
which could create new possibilities for model building.
Some preliminary investigations of this possibility have been carried out,
but the search for viable models remains largely an open problem.

The discussion in this section and the previous section shows
that when Lorentz symmetry is spontaneously broken
there will in general be both massless NG modes and
massive modes.
Clearly, any theory with spontaneous Lorentz breaking 
must account for these modes and what their role is
in the underlying dynamics described by the theory.
A more definite investigation along these lines
requires working in the context of a concrete model. 
A given model is defined by the rank of the tensor acquiring a vev
and by the forms of the kinetic and potential terms.
The simplest example is a theory with a vector field 
that has a nonzero vacuum value induced by a potential $V$.
A model of this type is known as a bumblebee model
\cite{ks,akgrav}.

\section{Bumblebee Models}

There are numerous examples of bumblebee models that
have been explored in recent years.
(For some specific examples, see
\cite{ks,akgrav,rbak,rbffak,kl01,baak05,kb06,
clay01,ejm,kt02,mof03,grip04,cl04,bp05,ems05,lr05,gs05,acfn,clmt06}).
They all involve a vector field $B_\mu$
that acquires a fixed vacuum value $b_\mu$.
They can be defined generally in Riemann-Cartan spacetime,
or restrictions to Riemann or Minkowski spacetime can be considered.
A complete definition depends on the choice of kinetic and
potential terms for $B_\mu$ and the gravitational fields.
A general Lagrangian typically has the form
\beq
\cL = \cL_0 - V(B_\mu B^\mu \pm b^2) + \cL_{\rm M} ,
\eeq
where $\cL_0$ contains the kinetic terms,
$V$ is the potential that induces spontaneous Lorentz breaking,
and $\cL_{\rm M}$ contains additional interaction and matter terms.
A particularly noteworthy feature that all bumblebee models share is that
they do not have a local $U(1)$ gauge symmetry.
This symmetry is broken explicitly by the form of the potential $V$,
which in general has a functional form involving products of $B_\mu$,

The specific choice of kinetic terms is largely a reflection of how
the bumblebee field $B_\mu$ is to be interpreted.
One approach is to view $B_\mu$ as the vector
in a vector-tensor theory of gravity.
In this case,
an appropriate kinetic term has a form similar
to that investigated by Will and Nordvedt
\cite{cmw},
\bea
\cL_0 &=& \fr 1 {16 \pi G} R 
+ \si_1 B^\mu B^\nu R_{\mu\nu}
+ \si_2 B^\mu B_\mu R
- \fr 1 4 \ta_1 B_{\mu\nu} B^{\mu\nu} 
\nonumber \\
&& \quad\quad\quad\quad\quad\quad\quad\quad\quad
+ \fr 1 2 \ta_2 D_\mu B_\nu D^\mu B^\nu
+ \fr 1 2 \tau_3 D_\mu B^\mu D_\nu B^\nu  ,
\eea
where $B_{\mu\nu} = D_\mu B_\nu - D_\nu B_\mu$.
A generalization of this form adds an additional 
fourth-order term in $B_\mu$
\cite{ejm}.
In this type of approach,
it is common to assume only gravitational couplings to matter, 
and therefore the terms $\cL_{\rm M}$ are not directly
relevant and can be dropped.

An alternative interpretation of the vector field $B_\mu$ is
that it is a generalized vector potential.
In this case,
the field strength 
$B_{\mu\nu}$ can be considered
the more physically relevant quantity,
and the natural choice of kinetic terms have an Einstein-Maxwell form,
as first considered by Kosteleck\'y and Samuel (KS)
\cite{ks},
\beq
\cL_0^{\rm KS}  = \fr 1 {16 \pi G} R - \fr 1 4 B_{\mu\nu} B^{\mu\nu} .
\label{LEM}
\eeq
Note that there is still no local $U(1)$ gauge symmetry
in this class of models when a nonzero potential $V$ is included
in the full Lagrangian.
However,
it is common in this case to include couplings to matter
along with some basic notion of charge in the matter sector.
For example,
terms involving current couplings with charged matter
can be included by defining,
$\cL_{\rm M} = B_\mu J^\mu$ with $D_\mu J^\mu = 0$.
In this case, the theory has a global $U(1)$ symmetry that
gives rise to charge conservation in the matter sector.

In a similar way,
there are different choices that can be made for the potential $V$.
One choice is a smooth quadratic potential,
\beq
V = \half \ka (B_\mu B^\mu \pm b^2)^2 ,
\label{Vkappa}
\eeq
where $\ka$ is a constant (of mass dimension zero).
This type of potential allows both NG excitations 
(obeying $V^\prime = 0$)
as well as massive excitations
(obeying $V^\prime \ne 0$).
A second example is a linear Lagrange-multiplier potential
\beq
V = \la (B_\mu B^\mu \pm b^2) , 
\label{Vsigma}
\eeq
where $\la$ is a Lagrange-multiplier field.
In this case,
the Lagrange multiplier field $\la$ imposes a constraint,
$B_\mu B^\mu = \mp b^2$,
which only allows NG excitations in $B_\mu$ and
excludes massive-mode excitations.
However,
there is still an additional degree of freedom
in the form of the Lagrange-multiplier field $\la$,
and its effects on dynamics must be understood along with those 
due to the NG modes.

\subsection{KS Bumblebee Model}

To illustrate the behavior of the NG and massive modes
and for definiteness,
consider the case of a KS bumblebee model.
In the absence of a cosmological-constant term,
it has a Lagrangian with the kinetic term 
$\cL_0^{\rm KS} $ in \rf{LEM}.
To allow for effects due to a massive mode,
the potential $V$ is chosen as the 
smooth quadratic potential in \rf{Vkappa}.
For simplicity,
the vacuum value $b_\mu$ is taken as timelike,
and an interaction of the form $B_\mu J^\mu$ is chosen.
To begin the analysis,
a Riemannian spacetime geometry is assumed.
Generalization to a Riemann-Cartan geometry is deferred to 
a later section.

The equations of motion for the KS bumblebee model
are obtained by varying the Lagrangian
with respect to the metric and bumblebee fields.
The results are
\beq
G^{\mu\nu} = 8 \pi G T^{\mu\nu} ,
\label {geq}
\eeq
\beq
D_\nu B^{\mu\nu} =  J^\mu - 2 V^\prime B^\mu  .
\label{Beq}
\eeq
Here, $G^{\mu\nu}$ is the Einstein tensor
and $T^{\mu\nu}$ is the total energy-momemtum tensor,
which consists of two terms,
\beq
T^{\mu\nu} = T^{\mu\nu}_M + T^{\mu\nu}_B .
\label{Ttotal}
\eeq
$T^{\mu\nu}_M$ is the energy-momentum tensor for the matter sector,
while the bumblebee energy-momentum is given by
\beq
T^{\mu\nu}_B = B^{\mu\al} B^\nu_{\pt{\nu}\al} - \fr 1 4 g^{\mu\nu} B_{\al\be} B^{\al\be}
- V g^{\mu\nu} + 2 V^\prime B^\mu B^\nu . \quad
\label{TBmunu}
\eeq
The contracted Bianchi identities for $G_{\mu\nu}$
lead to conservation of the total energy-momentum tensor,
\beq
D_\mu T^{\mu\nu} = D_\mu (T^{\mu\nu}_M + T^{\mu\nu}_B) = 0 .
\label{Tconsv}
\eeq
Similarly,
the antisymmetry of the bumblebee field strength $B_{\mu\nu}$
leads to a current-conservation law following from \rf{Beq},
\beq
D_\mu (J^\mu - 2 V^\prime B^\mu) = 0 .
\label{Jconsv}
\eeq
Examination of these equations reveals 
that when a massive mode is present,
with $V^\prime \ne 0$,
it acts as both a source of energy and charge density.
However,
in the absence of a massive mode,
$V^\prime = V = 0$,
and the equations reduce to the usual Einstein-Maxwell equations.
Since the NG modes obey the condition $V^\prime = 0$,
these modes by themselves obey the same equations
as in electrodynamics in a gravitational background.
This raises the interesting possibility that massless photons
arise in this type of model not as a result of gauge invariance
but instead as a result of the appearance of NG modes when
Lorentz symmetry is spontaneously broken.  

The idea that photons might arise as NG modes
due to spontaneous Lorentz breaking arose initially
in the context of special relativity,
where Lorentz symmetry is a global symmetry.
In these early models,
e.g., the model of Nambu
\cite{nambu68},
the nonzero vacuum value is imposed as a nonlinear
gauge choice in the context of a theory with local U(1)
gauge invariance.
As a result of this,
there are no physical signatures of Lorentz violation.
The bumblebee models are different in that they do not have
local U(1) gauge invariance.
They also permit matter couplings with the vacuum value $b_\mu$,
which can provide physical signatures of Lorentz violation.

A more complete determination of whether Einstein-Maxwell
solutions can emerge from bumblebee models,
requires understanding the role of the massive mode.
It constitutes an additional degree of freedom beyond those
of the NG modes that must be accounted for.
It also alters the form of the initial-value problem.
However, an exact solution of the equations of motion is not feasible,
since they are highly nonlinear in form.
In particular,
the massive mode couples nonlinearly to both the NG and
gravitational modes by acting as a 
source of charge and energy density.
Moreover,
with this extra degree of freedom present,
the full nonlinear theory is known to have at least
one allowed initial value (in a Minkowski-spacetime limit)
for which the Hamiltonian is negative and unbounded from below
\cite{clay01,ejm}.
It is possible, however, to restrict the theory in such a way
that the Hamiltonian remains positive.
One assumption is that the matter current $J^\mu$ is
conserved and does not mix with the effective bumblebee 
charge stemming from the nonlinear field interactions.
As stated above,
this requires that the matter sector has a conserved charge
(as expected in ordinary matter),
and therefore the theory has a global U(1) symmetry.
With this assumption,
initial values can then be chosen that separate
the full phase space into regions that do not mix.
In particular,
a region of phase space that maintains a positive
Hamiltonian can be selected.

It may be possible as well to alter the stability of the theory
by adding nonrenormalizable terms to the potential $V$.
It has, for example, been shown (in Minkowski spacetime)
that nonpolynomial potentials can lead to spontaneous Lorentz breaking
and that such potentials are stable
\cite{baak05}.
Ultimately, however,
the potential instability is not likely to be relevant for physics.
Viewing the bumblebee as an effective theory arising from
a more fundamental and stable quantum theory of gravity,
the apparent instabilities would merely reflect an incomplete 
knowledge of the physics entering at energy scales above 
that of the effective theory.
However,
in the absence of a fundamental quantum theory,
it is not possible to pursue these questions further.

For the purposes considered here,
with the aim of illustrating the effects of the NG and massive modes
in a gravitational theory,
it suffices to consider the bumblebee model
with a KS kinetic term and conserved matter currents.
It also suffices to work in the linearized limit.
In such a limit,
the Hamiltonian is positive (in a Minkowski-spacetime limit),
while the massive mode retains its feature of behaving as
a source of both charge and energy density.
Hence,
this limit is suitable for examining the effects 
of the massive mode on the gravitational interactions.

\subsection{NG and Massive Modes}

Solutions for the diffeomorphism and Lorentz NG modes 
can be obtained directly in the linearized approximation.
With a vector vev $b_\mu$,
symmetry under
three Lorentz transformations and one diffeomorphism
are spontaneously broken.
Thus, there can be up to three Lorentz NG modes and
one diffeomorphism NG mode.
Using a vierbein formalism,
the NG modes can be written as small virtual transformations away from
the vacuum solution.
Or, alternatively, 
gauge choices can be made that leave
the NG modes as combinations of the bumblebee excitations
$\cE^\mu = (B^\mu - b^\mu)$ and the metric excitations $h_{\mu\nu}$.

First, considering the diffeomorphism NG mode,
it is found that it drops out completely from the linearized theory
and does not propagate as a physical massless mode
\cite{rbak}.
Indeed, any Lagrangian formed out of contractions of
the curvature tensor and the field strength $B_{\mu\nu}$
will not contain an NG mode for the broken diffeomorphisms.

In contrast,
the Lorentz NG modes are found to consist of
two propagating transverse massless modes
and one auxiliary mode that does not propagate
\cite{rbak}.
They obey a condition that can be written in terms of
$\cE^\mu$ and $h_{\mu\nu}$ as
$b^\mu (\cE_\mu - \half h_{\mu\nu} b^\nu) = 0$,
which resembles a type of axial-gauge 
condition in electromagnetism in the presence of gravity.
Hence,
as expected, 
it is found that the Lorentz NG modes behave like photons
in curved spacetime.
However,
it must be stressed again 
that the bumblebee models in general have
additional matter couplings that can provide physical
signatures of Lorentz violation,
so the NG sector coupled to matter is not strictly speaking 
equivalent to Einstein-Maxwell theory.

A massive mode consisting of field excitations
that do not stay in the potential minimum can
also occur as a solution of the equations of motion
\cite{rbffak}.
Unlike the NG modes,
it cannot be written in terms of the vierbein and
vacuum values alone,
since the condition $B_\mu B^\mu = \mp b^2$ does
not hold for the massive mode.
At linear order, 
and in terms of $\cE^\mu$ and $h_{\mu\nu}$, 
the massive mode can be identified as the combination 
\beq
\be = \mp \fr {b^\mu (\cE_\mu - \half h_{\mu\nu} b^\nu)} {b^2} .
\eeq
It is clearly independent of the Lorentz NG modes,
which obey $\be = 0$.

To lowest order,
the equations of motion reveal a condition 
the massive mode must obey, 
\beq
b^\mu \partial_\mu \be \simeq 0 .
\label{betacond}
\eeq
For the case of a timelike vector field,
with $b_\mu = (b,0,0,0)$,
this condition 
shows that the massive mode does not propagate 
as a free field in the linearized limit.
Instead,
it is purely an auxiliary field $\be (\vec x)$ that has no time dependence.
As a result,
its value is fixed by the initial conditions at $t=0$.
Although it does not propagate,
it can nevertheless alter the form of the static potentials.

As an example of this,
consider a static point particle with mass $m$ and
charge $q$.
In the absence of Lorentz violation the static potentials
are the usual Coulomb potential $\Ph_q = q/{4 \pi r} $
and the Newtonian gravitational potential $\Ph_m = - {Gm}/r$.
Both $\Ph_q$ and $\Ph_m$ obey Poisson equations
that determine the form of these potentials.
Each has a source given by
the point-particle mass or charge density.
In the presence of spontaneous Lorentz violation,
it is convenient to introduce a third potential $\Ph_B$
for the bumblebee massive mode.
It is also defined by a Poisson equation,
\beq
\vec \nabla^2 \Ph_B (\vec x) = - \rh_B ,
\eeq
where it is the massive mode $\be (\vec x)$
that acts as a source of density
$\rh_B = - 4 \ka b^2 \be$.
It is this extra degree of freedom that enters in the equations
of motion and alters the form of the electromagnetic
and gravitational static potentials.

Electric and magnetic fields can be defined with the usual form,
but as functions of the bumblebee excitations.
First define
$F_{\mu\nu} = \partial_\mu \cE_\nu - \partial_\nu \cE_\mu$,
and solve for its components using the linearized field equations.
The $\vec E$ and $\vec B$ fields can then be determined
for the case of a static point particle.
It is found that the fields are modified by the presence of the massive mode
and are given as
\beq
\vec E = - \vec \nabla \Ph_q - \vec \nabla \Ph_B ,
\quad\quad
\vec B = 0 .
\label{EB}
\eeq
Evidently,
there is no static magnetic field generated for the case of a
purely timelike vacuum value $b_\mu$.
However,
the static electric field is modified by the presence of the 
massive-mode potential.
Even for a neutral point mass (with $q=0$),
a nonzero massive mode can generate a
nonzero electric field.

Similarly,
the modified gravitational potential $\Ph_g$ can be
determined from the field equations of motion.
For the case of a point mass,
it is found to have the form
\beq
\Ph_g = \Ph_m - 4 \pi G b  \Ph_B .
\label{Phigsol}
\eeq
Clearly, the gravitational potential is altered
by the massive mode $\be (\vec x)$.
However,
the specific form that the potential takes depends on the choice
of the initial value for the massive mode.
This opens up the possibility of exploring modified forms
of the gravitational potential in search of,
for example, an alternative explanation of dark matter.
In fact,
there is considerable freedom in this approach,
since the only experimental constraints are that the
potential must agree with the usual Newtonian potential
over probed distance scales.

Given the lack of specific experimental guidance, 
a natural choice of initial value
would be to set $\be = 0$ at time $t=0$.
In this case,
$\Ph_B = 0$,
and the static potentials reduce to the usual Coulomb
and Newtonian expressions.
This holds true as well with a nonzero massive mode
if the scale $|M^2| = 4 \ka b^2$ becomes extremely large,
approaching the Planck scale, for example.
Here again,
the electromagnetic and gravitational
potentials approach their conventional values.
These results in particular reveal that the usual
Einstein-Maxwell solutions 
(describing massless photons
as well as the correct static potentials) 
can emerge from a theory that has no local U(1) gauge symmetry.
The photons in this case are the Lorentz NG modes,
and the massive mode remains extremely heavy and
thus has little effect on the static potentials.

Other examples with $\Ph_B \ne 0$ can be considered as well.
In these cases,
both the static gravitational and Coulomb potentials
are modified by the massive mode.
One simple example is the choice
$\Ph_B = - \Ph_q$.
It has $\Ph_g \ne \Ph_m$,
and hence has a modified Newtonian potential.
The solution of the bumblebee field has the form
of a total derivative
$\cE_\mu = \partial_\mu \ch$, 
where $\ch$ is a
scalar depending on $\Ph_m$ and $\Ph_q$
\cite{rbffak}.
The field strength $F_{\mu\nu}$ vanishes
because the bumblebee density cancels the
charge density $\rh_q$.
However,
the bumblebee excitations $\cE_\mu$ remain
nonzero through the dependence on $\Ph_m$.
This type of solution 
has an unusual behavior that has been
identified as potentially flawed \cite{clay01,ejm}
due to the formation of shock discontinuities in
$\cE_\mu$.
However,
in fact, 
this behavior is to be expected.
For a point charge,
the singularities in $\cE_\mu$ merely reflect the
fact that the field has $1/r$ dependence stemming from
its dependence on $\Ph_m$.
Indeed,
the same behavior appears in the usual
solutions of Einstein-Maxwell theory in an
appropriately chosen gauge.

\subsection{Higgs Mechanism for the Spin Connection}

In Riemann-Cartan spacetime,
the possibility of a Higgs mechanism involving the
Lorentz NG modes becomes a possibility.
In this mechanism it is the spin connection that
gains mass terms as a result of spontaneous Lorentz breaking.
As long as the spin connection is dynamical,
this mechanism can in principle be viable,
leading to physical massive propagating spin-connection fields.
In practice, however,
it is difficult to construct a model that is ghost- 
and tachyon-free.

An illustration is provided by the KS bumblebee
model in Riemann-Cartan spacetime.
In this case,
when $B_\mu$ has a vacuum value $b_\mu$,
the field strength $B_{\mu\nu}$ can be written in
terms of the vierbein and spin connection as
\beq
B_{\mu\nu} =
(\vb \mu \be  \lulsc \nu \al \be - \vb \nu \be  \lulsc \mu \al \be ) b_\al .
\label{FS}
\eeq
When $B_{\mu\nu}$ is squared,
quadratic terms in $\lulsc \mu \al \be$ appear in the lagrangian,
which perturbatively have the form
\beq
- \quar e B_{\mu\nu} B^{\mu\nu} 
\approx 
- \quar (\om_{\mu\rh\nu} - \om_{\nu\rh\mu})
(\om^{\mu\si\nu} - \om^{\nu\si\mu}) b^\rh b_\si .
\label{om2}
\eeq
It is these quadratic terms 
that allow a Higgs mechanism to occur
involving absorption of the Lorentz NG modes 
by the spin connection.

In Ref.\ \cite{rbak},
a number of models with generalized kinetic terms 
for the spin connection were considered.
Finding a physical model with no ghosts, however,
remains an open problem.
The difficulty is in finding kinetic terms describing propagating modes
that are compatible with Eq.\ \rf{om2} as a mass term.
If ghosts are permitted,
then the mechanism is straightforward.
For example, with a kinetic term in the gravitational
sector of the form
\beq
\cl_{0,{\rm grav}} =  
\frac 1 4 R_{\la \ka \mu \nu} R^{\la \ka \mu \nu}  .
\label{R2L}
\eeq
all the fields $\lsc \la \mu \nu$ with $\la \ne 0$
propagate as massless modes.
When this is combined with the mass term \rf{om2},
some of the propagating modes 
are converted to massive modes.
Other examples can be studied as well,
aided by decomposing the fields $\lsc \la \mu \nu$
according to their spin-parity projections.
Evidently, the incorporation of spontaneous Lorentz violation 
in theories with torsion provides a new arena in the 
search for models with propagating massive modes.
The challenge, however, is to find viable models that
do not allow ghosts.

\section{Phenomenology of Lorentz Violation}

The effects described in this paper originating from
spontaneous Lorentz breaking can provide clear signals 
of physical Lorentz violation.
In the end,
there are basically three classes of signals:
those arising from NG modes, from massive modes, and
from matter couplings.
The phenomenological implications of each of these
can be considered.

The NG modes either lead to additional gravitational
modes in the vierbein that would differ from the usual
forms of gravitational radiation predicted in general relativity,
or in certain cases they can be interpreted as known gauge
fields such as the photon or graviton
\cite{kp05}.
However,
in the latter case,
there would be no observable consequences of the NG modes
themselves (at least at leading order)
other than the existence of the previously known massless gauge particles.

Massive modes can arise in two ways,
either through a Higgs mechanism in Riemann-Cartan spacetime,
or as field excitations that do not remain in the potential minumum.
In either case,
they could in principle be detectable as new
previously unobserved propagating particles.
Alternatively, however,
the massive modes might remain auxiliary fields that do not propagate,
as in the example of the KS bumblebee.
Their influence then appears to be
limited to altering the form of the relevant static potentials.
In either of these scenarios,
it is likely that the scale associated with the massive modes is
extremely high,
and therefore their observable consequences are likely to be quite small.

Any remaining signals of spontaneous Lorentz violation
would involve couplings with matter fields.
For example,
at low energy,
signals of physical Lorentz violation would occur
when a tensor vev, e.g., $\vev{T_{\la\mu\nu}}$, 
couples with
conventional Standard-Model or gravitational fields.
As a result,
any possible signal originating in this way would be
identical to a signal arising in the SME.
This is because
the SME allows for all observer-independent violations of Lorentz symmetry 
involving Standard-Model and gravitational fields.
It is defined as a general effective field theory at low energy,
but it also provides a connection to the Planck scale 
through operators of nonrenormalizable dimension 
\cite{kl01}.
In many cases, it is sufficient to restrict the full SME to
minimal extensions involving only, for example,
power-counting renormalizable or SU(3)$\times$S(2)$\times$U(1)
gauge-invariant terms.
To consider experiments in atomic physics it often suffices
to restrict the SME to its QED sector.
Similarly,
limits of the SME that include (or exclude) gravity can be defined,
and in the case with gravity either a Riemann or Riemann-Cartan geometry
can be assumed.

A group of theorists centered at Indiana University initiated a
comprehensive phenomenological investigation of Lorentz
violation more than a decade ago.
These investigations span virtually all areas of physics.
The scope of these investigations includes searching for
signals of spontaneous Lorentz violation
(as well as other, e.g., explicit, forms of Lorentz breaking).
This work has stimulated a number of new and improved
experiments.
These include classic tests of Lorentz and CPT symmetry,
such as $g-2$ experiments in Penning traps,
Hughes-Drever experiments,
modern-day Michelson-Morley experiments,
as well new types of tests,
such as in space satellites or with
astrophysical sources.
Together they cover a wide range of particle sectors
in the Standard Model.
Some specific examples 
include tests with
electrons
\cite{electrons1,electrons2},
muons
\cite{muons},
hadrons
\cite{cc,hadrons},
neutrinos
\cite{neutrinos},
and photons
\cite{photons}.
Many of these efforts are on-going,
with plans for attaining significantly
improved sensitivities in the coming years.

\section{Summary and Conclusions}

This work has examined possible consequences of
spontaneous Lorentz violation in the context of gravity.
Much of the focus has been on questions concerning
the fate of the NG modes, the possibility of a Higgs
mechanism, and the appearance of additional massive modes.
In general, it is found that 
in theories with spontaneous Lorentz violation, 
up to ten NG modes can appear. 
They can all be incorporated naturally in the vierbein.  
For the example of a KS bumblebee model, 
the Lorentz NG modes can propagate like photons in an axial gauge. 
In principle, two Higgs mechanisms can occur,
one associated with broken diffeomorphisms,
the other with Lorentz symmetry.
However,
it has been shown that a conventional Higgs mechanism (for diffeomorphisms)
involving the metric does not occur.
If the geometry is Riemann-Cartan, 
then a conventional Higgs mechanism (for the Lorentz symmetry)
can occur, 
in which the spin connection acquires a mass.
However,
in a Riemann geometry,
this type of Higgs mechanism is not possible.
Nonetheless,
an alternative type of Higgs mechanism can occur,
leading to the appearance of additional massive modes involving
the metric field.
These can lead to altered forms of the gravitational potential.
Clearly,
there are numerous phenomenological questions that arise in
these processes.
However,
all relevant signals of Lorentz breaking at low energies
invloving couplings to known Standard-Model fields can be
pursued comprehensively using the SME.

\section*{Acknowledgments}

This work has been supported by
NSF grant PHY-0554663 and by
the European Science Foundation network
programme ``Quantum Geometry and Quantum Gravity.''

\end{document}